\documentclass[aps,pra,preprint,superscriptaddress]{revtex4}
\usepackage{graphicx}% Include figure files
\usepackage{dcolumn}% Align table columns on decimal point
\usepackage{bm}% bold math
\usepackage{color}% couleur en rgb

\begin{document}

%\preprint{}

\title{Characterization and limits of a cold atom Sagnac interferometer}

\author{A. Gauguet\footnote{Present address: Department of Physics, Durham University, Rochester Building, South Road, Durham DH1 3LE, England}}
\author{B. Canuel\footnote{Present address: European Gravitational Observatory, Via E. Amaldi, 56021 S. Stefano a Macerata - Cascina (PI), Italy}}
\author{T. L\'{e}v\`{e}que}
\author{W. Chaibi\footnote{Present address: ARTEMIS - Observatoire de Nice, Boulevard de l'Observatoire, 06305 Nice, France}}
\author{A. Landragin}
\email[]{arnaud.landragin@obspm.fr}
%\thanks{}
\altaffiliation{LNE-SYRTE, UMR 8630 CNRS, UPMC, Observatoire de
Paris} \affiliation{ LNE-SYRTE, Observatoire de Paris, CNRS, UPMC,
61 avenue de l'Observatoire, 75014 Paris, FRANCE}

\date{\today}

\begin{abstract}
We present the full evaluation of a cold atom gyroscope based on
atom interferometry. We have performed extensive studies to
determine the systematic errors, scale factor and sensitivity. We
demonstrate that the acceleration noise can be efficiently removed
from the rotation signal, allowing us to reach the fundamental
limit of the quantum projection noise for short term measurements.
The technical limits to the long term sensitivity and accuracy
have been identified, clearing the way for the next generation of
ultra-sensitive atom gyroscopes.
\end{abstract}
%\keywords{}

\maketitle
\section{INTRODUCTION}
Inertial sensors are of interest in both science and industry.
High precision sensors find scientific applications in the areas
of general relativity~\cite{will}, geophysics~\cite{stedmann} and
navigation~\cite{navigation}. In these fields, matter-wave
interferometry is promising since it is expected to be an
extremely sensitive probe for inertial forces~\cite{clauser}. In
1991, atom interferometry techniques were used in
proof-of-principle demonstrations to measure
rotations~\cite{Riehle91} and accelerations~\cite{Chu91}. The
first demonstrations of highly sensitive atomic gyroscopes using
thermal beams~\cite{Gustavson97,Pritchard97} were obtained in
1997, followed by sensors with sensitivities at the
state-of-the-art level~\cite{Gustavson2000,Kasevich2006}. For
practical applications, cold atom interferometry is of fundamental
interest thanks to its intrinsic stability and accuracy, as the
measurement of inertial forces is realized with respect to the
inertial frame of the free-falling atoms. The use of cold atoms
allows better control of atomic velocity and interaction time,
leading to a better accuracy in a much more compact
instrument~\cite{canuel2006,wu2007,wang2007,rasel2009}.

In this paper we present the full characterization of a gyroscope
based on atom interferometry, sensitive to the Sagnac effect.
Different parameters have been taken into account for the study:
the short term noise, the stability of the systematic effects, the
scale factor and its linearity. The apparatus uses Cesium atoms
and Raman transition to manipulate the matter wave-packets. In our
setup, we use a single Raman beam interacting with slow atoms,
which makes the setup very versatile. Thus the experiment enables
us to measure the full basis of inertia (three components of
acceleration and rotation) with the same
apparatus~\cite{canuel2006}, making it suitable for applications
such as inertial navigation. In this paper, we emphasize the
possibility to measure the systematic effects and the scale factor
accurately thanks to our polyvalent apparatus. In
section~\ref{Apparatus} we describe the experimental setup and the
measurement process. A detailed analysis of the different sources
of systematic errors and tests of the scaling factor are presented
in section~\ref{Bias}. Finally, the analysis of the stability of
the rotation measurement and its main limitations are described in
section~\ref{Sensitivity}.
\begin{figure} \includegraphics[width=8cm,angle=0]{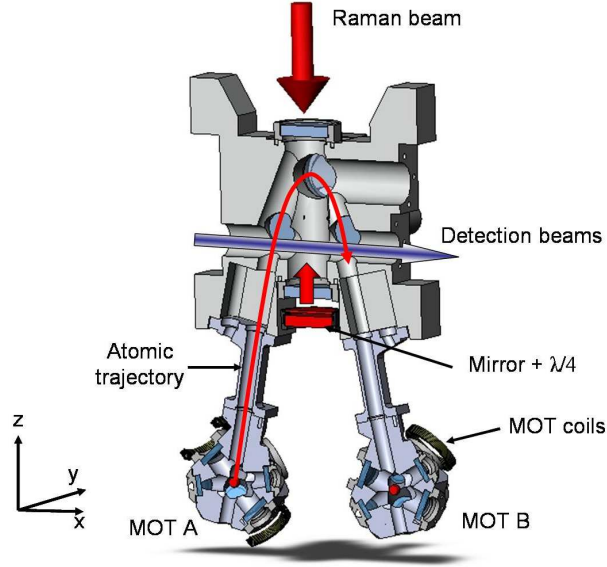}
\caption{(Color online) Scheme of the vacuum chamber showing the
two MOTs, the interferometer zone and detection probe. The total
dimensions of the system are 30~x~10~x~50~cm.} \label{setup}
\end{figure}

\section{Apparatus}\label{Apparatus}

A global view of the experimental setup is presented in
Fig.~\ref{setup}. The whole experiment is surrounded by two layers
of mu-metal shield in order to reduce the impact of external
magnetic fields. Cesium atoms are first loaded from a vapor into
two independent Magneto-Optical Traps (MOT), called A and B in the
following. The two Cesium clouds are then launched into two
opposite parabolic trajectories using the moving molasses
technique. At the top of their trajectory, the atoms interact
successively three times with a unique pair of retro-reflected
Raman beams, which acts on matter-waves as beam splitters or
mirrors. This creates an interferometer of 80~ms total maximum
interaction time. The atomic phase shift is then obtained from the
population in each output port, which is measured by a
fluorescence technique thanks to the state labelling of the
interferometer output ports~\cite{borde1989}.

In this paper we will focus on the configuration based on vertical
Raman beams. The use of two atomic sources allows us to
discriminate between the acceleration along the vertical direction
and the rotation around the y horizontal axis. The experiment is
mounted on a horizontal rotating platform, which enables us to
vary the projection of the Earth's rotation rate along the
sensitive axis of the gyroscope.

\subsection{Atomic preparation}

Cesium atoms are loaded from a thermal vapor during 140 ms into
two independent MOTs. After the MOT-coils are turned off, the
atoms are kept in an optical molasses for 15 ms, to allow the
stray magnetic field to decay. A frequency shift between the upper
and the lower cooling beams is then applied, to launch the atoms
thanks to the moving molasses technique. The independent control
of the lower and upper cooling beam frequencies is achieved by
passing through two different acousto-optic modulators (AOM). By
applying a frequency shift of 3.2~MHz on the AOM controlling the
lower cooling beams, the atoms are launched with a velocity of
2.4~m~s$^{-1}$ at an angle of 8$^{\circ}$ with respect to the
vertical direction. In addition, the atoms are cooled down to a
temperature of about 1.2 $\mu$K in the molasses by chirping the
mean frequency down to $-15 \Gamma$, with respect to the closed
transition ${|F=4\rangle \leftrightarrow |F'=5\rangle}$.

Following this launching stage, the atoms are distributed among
all Zeeman sub-levels of the ${|6S^{1/2},F=4\rangle}$ state. In
order to reduce the sensitivity to parasitic magnetic fields,
atoms are selected in the sub-level $m_F=0$. For this purpose, a
static magnetic field of 30~mG is applied in the z direction to
lift the degeneracy of the Zeeman sub-levels. Atoms in
${|6S^{1/2},F=4,m_F=0\rangle}$ are transferred to
${|6S^{1/2},F=3,m_F=0\rangle}$ when passing through a micro-wave
cavity. Any atoms remaining in ${|6S^{1/2},F=4\rangle}$ are then
removed by means of a pusher beam. After this preparation stage,
we obtain typically $10^7$ atoms in the
$|6S^{1/2},F=3,m_F=0\rangle$ ground state with a residual in the
other states of less than 1\%, for both sources.

\subsection{Implementation of the interferometer}
\subsubsection{Three pulse interferometer}
\begin{figure}
\includegraphics[width=7cm]{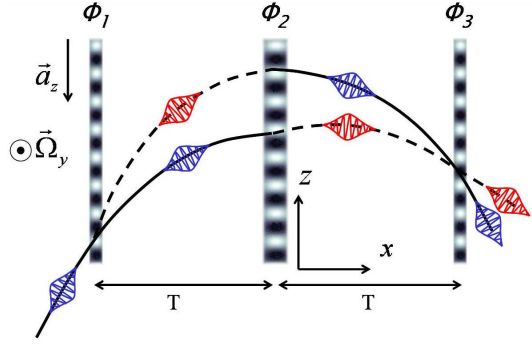}
\caption{(Color online) Scheme of the $\pi/2-\pi-\pi/2$
interferometer in the vertical Raman configuration with a total
interaction time of $2T$. The interferometer is sensitive to the
vertical acceleration~$a_z$ and the rotation~$\Omega_y$. Solid and
dotted lines respectively represent the partial wave packets in
the states $|6S_{1/2},F=3\rangle$ and $|6S_{1/2},F=4\rangle$.}
\label{interferometre}
\end{figure}

When the atoms arrive close to the apex of the parabolic
trajectories, occurring at $t_{ap}=244$~ms after launch, the wave
packets are split, deflected and recombined by stimulated Raman
transitions \cite{Chu91} in order to realize the interferometer.
Since the Raman beams are vertically oriented the interferometric
area is created in the (x,z) plane~(Fig.~\ref{interferometre}).

The output phase shifts $\Delta\Phi^{A}$ and $\Delta\Phi^{B}$ of
the two interferometers $\Delta\Phi^{A,B}$ are composed of three
terms which depend respectively on the acceleration~$\mathbf{a}$,
the rotation~$\mathbf{\Omega}$ and the Raman laser phase
differences of the three pulses~\cite{antoine}:
\begin{equation}
\label{delta_phi} \Delta\Phi^{A,B}=\Delta\Phi_{a} +
\Delta\Phi_{\Omega}^{A,B} + \Delta\Phi_{laser}
\end{equation}
In the vertical Raman configuration studied here, these
contributions are written as a function of the vertical
acceleration~$a_z$, the horizontal rotation~$\Omega_y$ and the
Raman laser phase differences~$\phi_{i}$, $i=1, 2, 3$:
\begin{eqnarray}
\label{facteur_echelle_equation}
\begin{array}{l l l}
\Delta\Phi_{a} & = &
k_\mathrm{eff}a_{z}\ T^{2}\\[0.5cm]
\Delta\Phi_{\Omega}^{A,B} & = &
2\ k_\mathrm{eff} \; V_{x}^{A,B} \Omega_{y}\;T^{2}\\[0.5cm]
\Delta\Phi_{laser} & = & \phi_{1}-2\phi_{2}+\phi_{3}
\end{array}
\end{eqnarray}

where $\mathbf{k}_\mathrm{eff}$ is the effective wave-vector of
the Raman beam. The rotation phase
shifts~$\Delta\Phi_{\Omega}^{A,B}$ measured by the two
interferometers are related to the horizontal velocities
$V_{x}^{A,B}$ and have opposite signs for the two sources. The use
of two counter-propagating sources allows to discriminate between
the acceleration and rotation phase shifts~\cite{Gustavson2000}.

\subsubsection{Raman lasers}
In order to drive Raman transitions between
$|6S_{1/2},F=3,m_F=0\rangle$ and $|6S_{1/2},F=4,m_F=~0\rangle$,
two counter-propagating laser beams, with a frequency difference
of~9.192~GHz, are required. These two optical frequencies are
generated by two extended cavity laser diodes~\cite{baillard06}
emitting at $\lambda=852$~nm. The first laser is locked by
frequency comparison to the MOT repumper beam with a detuning of
350~MHz with respect to the ${|6S_{1/2},F=3,m_F=0\rangle
\leftrightarrow |6P_{3/2}\rangle}$ transition. A second laser is
phase-locked to the previous one by comparing the beat note
between the two beams with a microwave reference~\cite{muwave}. In
order to get sufficient power to drive the transitions, the two
laser beams are injected into a common semiconductor tapered
amplifier~(EYP-TPA-0850-01000-3006-CMT03)~\cite{Leveque2009}. The
power ratio between the two lasers is adjusted close to 0.5 in
order to cancel the effect of the AC Stark shift~(see part~\ref{AC
stark shift}). This ratio is then finely tuned by means of Raman
spectroscopy on the cold atom samples. After amplification, the
two lasers have the same polarization and are guided to the atoms
through the same polarizing fibre. At the output of the fiber, the
beam is collimated with an achromatic doublet lens of 240~mm focal
length, giving a diameter at $1/e^2$ of~35~mm. The intensity at
the center of the beam is 20~mW~cm$^{-2}$. The two
counter-propagating beams are obtained thanks to a retro-reflected
configuration in which the two frequency beam passes through the
vacuum chamber and is reflected by a mirror, crossing a
quarter-wave plate twice~(Fig.~\ref{setup}). The quarter-wave
plate is set in such a way that retro-reflected polarizations are
orthogonal with respect to the incident ones. In this manner,
counter-propagating Raman transitions are allowed while
co-propagating ones are forbidden. This retro-reflected
configuration limits the parasitic effects induced by the
wave-front distortions, which are critical in order to achieve
good accuracy and long term stability~(see part~\ref{Wave-front
distortion}).

Moreover, since the atoms are in free fall, the frequency
difference between the two atomic levels is Doppler shifted in the
vertical direction by ${\omega_{D}=\mathbf{k}_\mathrm{eff}\cdot
\mathbf{g}(t-t_{ap})}$, which depends on the gravity $g$. In order
to satisfy the resonance condition during the whole atomic flight,
the frequency difference between the two Raman lasers is chirped
thanks to a Direct Digital Synthesizer~(DDS). Additionally the
interferometers are realized with a delay of 5~ms with respect to
the apex of the trajectories in order to avoid a null Doppler
shift during the $\pi$ pulse.

\subsection{Detection system}
\subsubsection{Detection apparatus}

With Raman transitions, state labelling~\cite{borde1989} enables
one to determine the momentum state of the atoms by measuring
their internal state. Thus, the phase shift of the interferometer
can be obtained simply by measuring the transition probability of
the atoms between the two ground states ${|6S_{1/2},F=3,m_F =
0\rangle}$ and ${|6S_{1/2},F=4,m_F = 0\rangle}$ at the output of
the interferometer. The population measurement is performed by
counting the number of fluorescence photons emitted by the atoms
at the crossing of successive resonant laser beams. These beams
are shaped to form horizontal slices of light with a rectangular
section of 10x5~mm so as to probe the whole atomic clouds.

\begin{figure}
\includegraphics[width=9cm,angle=0]{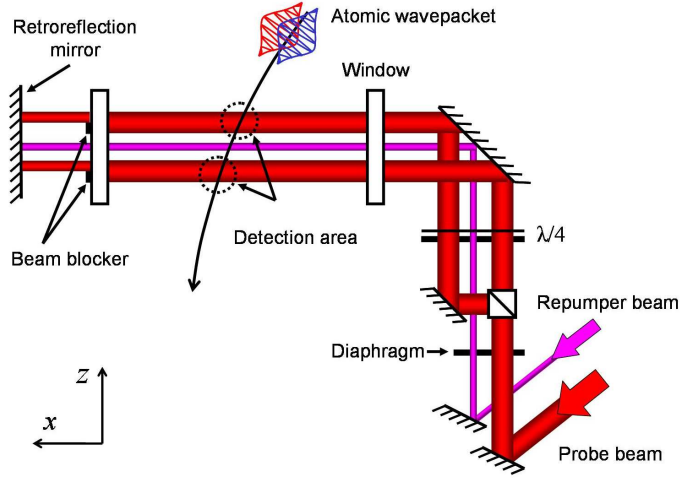}
\caption{(Color online) Scheme of the detection system. Two
retro-reflected probe beams are generated from the same laser
while a repumper beam is inserted between them. These three beams
are shaped to form horizontal slices using two slits.}
\label{schem-detection}
\end{figure}

The detection system is made up of three counter propagating beams
circularly polarized as shown in Fig.~\ref{schem-detection}. The
upper and lower beams are used as probes with a saturation
parameter of~0.3. They are obtained from the same laser, tuned to
the closed atomic transition ${|6S_{1/2},F=~4\rangle
\leftrightarrow |6P_{3/2},F'=5\rangle}$. A repumper beam is
inserted between these two probes and tuned to the transition
${|6S_{1/2},F=3\rangle \leftrightarrow |6P_{3/2},F'=4\rangle}$.

After the interferometric sequence, the atoms pass through the
first probe. The fluorescence light emitted by the cycling
transition ${|6S_{1/2},F=~4\rangle \leftrightarrow
|6P_{3/2},F'=5\rangle}$ is collected by an imaging system~(3.7
$\%$ efficiency) and focused onto a photodiode~(Hamamatsu 1327BQ).
The output signal $s_{1}$, given by the area of the time-of-flight
signal, is proportional to the number of atoms projected into the
$|6S_{1/2},F=4\rangle$ state. The lower part of this first probe
beam is not retro-reflected in order to eliminate these atoms. The
remaining cloud, made up of atoms in the $|6S_{1/2},F=3\rangle$
ground state, is optically pumped to the $|6S_{1/2},F=4\rangle$
state while passing through the repumper beam. These atoms are
finally detected with the second probe beam, providing a signal
$s_{2}$ proportional to the number of atoms initially in the state
$|6S_{1/2},F=3\rangle$ at the output of the interferometer. Each
interferometer uses two imaging systems in order to collect the
signal emitted by the atoms in the two respective ground states.
The transition probability $P$ is deduced from the fluorescence
signals obtained on the two photodiodes ($s_{1}$ and $s_{2}$) and
is written: $ P = \frac{s_{1}}{s_{1} + s_{2}}$.

\subsubsection{Detection noise analysis}
\label{detection} The noise affecting the fluorescence signals can
be separated into three main contributions~\cite{Santarelli1999}.
The first is related to power or frequency fluctuations of the
probe beams that induce a noise which scales linearly with the
total number of atoms $N^{A}$ and $N^{B}$ in each source. The
second contribution consists of a technical noise related to the
detection system (photodiode dark current and amplifier noise)
that gives a contribution independent of the total number of atoms
in the probe. Finally, the quantum projection noise~(QPN) gives a
fundamental limit of the measurement~\cite{Itano1993} and scales
as~$\left(P^{A,B}(1-P^{A,B})N^{A,B}\right)^{1/2}$. These
independent sources of noise can be added quadratically, giving a
variance of the transition probability:
\begin{equation}
\sigma_{P^{A,B}}^2 = \alpha + \frac{P^{A,B}(1-P^{A,B})}{N^{A,B}} +
\frac{\gamma}{{N^{A,B}}^2} \label{sigmaP}
\end{equation}

In order to determine the parameters $\alpha$ and $\gamma$ when
working at one side of a fringe of the interferometer, we use a
single Raman laser pulse giving an average transition probability
close to 0.5. In such a case, the noise detection properties are
similar to those of the interferometer without being sensitive to
the interferometer phase noise. In practice, the Raman laser power
is reduced so that the Raman pulse duration $\tau =
\pi/\Omega_{Rabi}$ limits the average transition probability to
$P^{A,B} =$ 0.5 by velocity selection. By this means, the
measurement is made almost insensitive to any power fluctuations
of the Raman beam.
\begin{figure}
\includegraphics[width=8cm,angle=0]{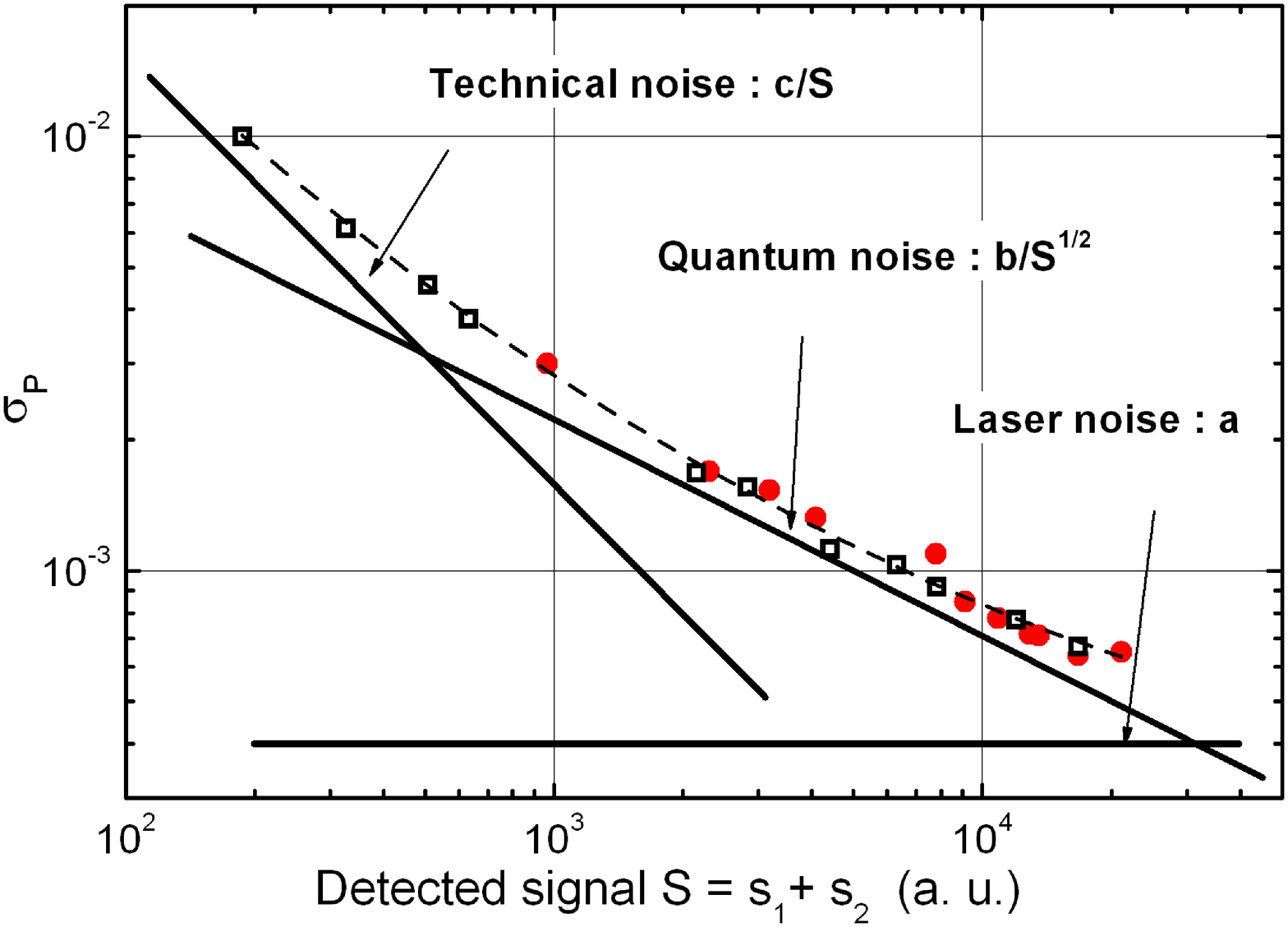}
\caption{(Color online) Detection noise analysis. The shot-to-shot
Allan standard deviation of the transition probability is plotted
versus the detected signal $S$ in the same arbitrary unit for the
two sources (respectively black squares and red dots). The dashed
curve corresponds to the fit of the data by the noise model, which
can be separated into three terms corresponding to the three solid
lines.} \label{detection_bruit}
\end{figure}
\begin{table}[htbp]
  \centering
\begin{tabular}{ |ll l  |}
    \hline
    Noise & Fitted parameters  & \\ \hline \hline
    Laser & $a = \sqrt{\alpha}$ &$ =4\;10^{-4}$\\
    QPN& $b = 1/(2\sqrt{\eta})$  &$ = 7\;10^{-2}$\\
    Technical & $c = \sqrt{\gamma}/\eta$  &$ = 1.6$\\ \hline
  \end{tabular}
  \caption{Table of the fitted parameters deduced from the measurement of the detection noise, presented in Fig.~\ref{detection_bruit}.}
  \label{parametersFit}
\end{table}

The detection noise $\sigma_{P}$ is evaluated as a function of the
number of detected atoms, which is changed by varying the loading
time of the MOTs (see~Fig.~\ref{detection_bruit}). The data are
scaled as a function of the fluorescence signal $S=s_{1}+s_{2}$,
which is proportional to the actual atom number $N^{A,B}=\eta S$.
The two detection systems clearly show identical noises for the
two atomic sources. A fit of the experimental data has been
realized by eq.~\ref{sigmaP} in which the scaling of the number of
atoms is a free parameter. This leads to a fit with three
parameters ($a$, $b$, $c$), related to $\alpha$, $\gamma$ and
$\eta$, which are reported in Table~\ref{parametersFit}. For the
usual experimental parameters, the number of atoms detected
corresponds to $S = 7000$ and the signal to noise ratio is then
limited by the quantum detection noise. In this case, the
parameter $b$ allow the determination of the actual number of atom
detected for each source, giving $N=3.6 \times 10^5$~atoms, which
is in agreement with additional measurements realized by
absorption.

\subsection{Measurement process}\label{Interf}
\begin{figure}
\includegraphics[width=8cm]{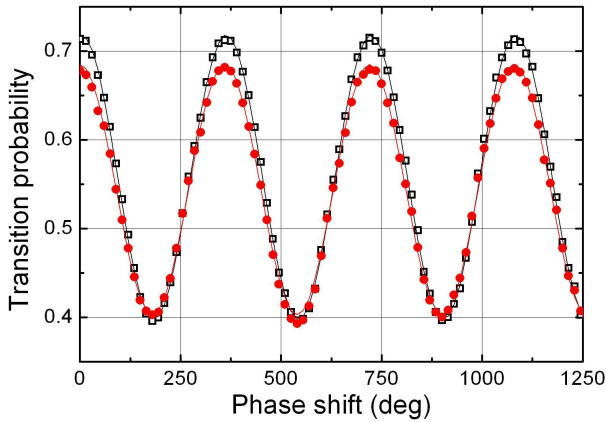}
\caption{(Color online) Atomic fringe patterns obtained for the
interferometers A~(black squares) and B~(red dots) with an
interaction time of $2 T=80$~ms and a pulse duration of 13~$\mu$s.
In order to scan the fringe patterns we incremented the effective
Raman laser phase at each measurement cycle.} \label{fig-franges}
\end{figure}

Interference fringe patterns are scanned by taking advantage of
the Raman laser phase control. Indeed, by adding a laser phase
offset $\delta\varphi_n$ between the second and the third pulses
of the $n^{th}$ measurement, the atomic phase measured evolves as
$\Delta \Phi_{laser} = \phi_1-2\phi_2+(\phi_3+\delta\varphi_n)$.
In practice, the phase increment $\delta\varphi_n$ is applied to
the micro-wave signal used as a reference to phase-lock the two
Raman lasers. The fringe patterns, shown in
Fig.~\ref{fig-franges}, were obtained with the usual parameters,
$2T=80$~ms and Raman pulse durations of ${\tau = 13\; \mu}$s. Both
atomic sources exhibit a contrast $C$ close to~30\% mainly limited
by the Raman transition efficiency. Two phenomena are responsible
for this limitation. On the one hand, the atomic cloud velocity
distribution is broader than the velocity linewidth of the Raman
transition. On the other hand, because of the thermal expansion of
the atomic cloud, all the atoms do not experience the same laser
power in the gaussian Raman beam, and therefore do not experience
the same Rabi frequency.

The transition probability measured at the output of the
interferometer is written as a function of the following terms: an
offset $M^{A,B}$, a contrast $C^{A,B}$, a phase shift due to
inertial effects~$\Delta \Phi_{I}^{A,B}$ (acceleration and
rotation) and a laser phase shift~$\Delta \Phi_{laser}$:
\begin{eqnarray}
\begin{array}{l}
P^A = M^A+C^A\cos(\Delta\Phi_{I}^{A}+\Delta \Phi_{laser})\\[0.5cm]
P^B = M^B+C^B\cos(\Delta\Phi_{I}^{B}+\Delta \Phi_{laser})
\end{array}
\label{probaAB}
\end{eqnarray}

The phase shift accumulated by the two interferometers can then be
deduced by fitting each fringe pattern with the
equation~\ref{probaAB}. The acceleration and rotation phase shifts
are then discriminated by calculating respectively the half sum
and half difference of the two fitted phase shifts:
\begin{eqnarray}
\begin{array}{l c r}
\Delta\Phi_{a} = \frac{\Delta\Phi^{A}+\Delta\Phi^{B}}{2} & \; &
\Delta\Phi_{\Omega} = \frac{\Delta\Phi^{A}-\Delta\Phi^{B}}{2}
\end{array}
\end{eqnarray}
In the general case, the interferometric phase shift can always be
extracted from four points: ${\Delta\Phi_{laser} = 0^\circ,
\;+90^\circ,\;+180^\circ \; \mathrm{and}\;270^\circ}$.

With the retro-reflected configuration, atoms are submitted to
four laser waves which couple Raman transitions along two opposite
effective wave-vectors~$\pm \mathbf{k}_\mathrm{eff}$. The opposite
Doppler shift between these two effective transitions allows the
deflection of the atomic wave packets along one or the other
direction by changing the sign of the frequency ramp delivered by
the Direct Digital Synthesizer. The sign of the inertial phase
shifts $\Delta\Phi_{I}^{A,B}$ changes according to $\pm
\mathbf{k}_\mathrm{eff}$. Consequently, by processing the half
difference of the phase shift measured at $\pm
\mathbf{k}_\mathrm{eff}$, parasitic phase shifts independent of
the direction of the effective wave vector are removed thanks to
this k-reversal technique.

To sum up, the experimental sequence consists in acquiring
transition probability measurements on the two sources,
alternately for each direction of the wave vector $\pm
\mathbf{k}_\mathrm{eff}$. The fit is processed afterwards and we
infer a measurement of the inertial phase shifts for each set of
eight acquisition points.

\section{Systematic errors and scale factor}\label{Bias}

\subsection{Characterization of the atomic trajectories}

A crucial point, when measuring a differential phase shift between
two
interferometers~\cite{Gustavson2000,mcguirk2002,lamporesi2008}, is
the overlap of the two atomic trajectories. Indeed, if the
trajectories are perfectly overlapped, many systematic effects
cancel out, as explained in paragraph~\ref{Systematic errors
estimation}. Consequently, the two cold atom sources were realized
with particular care. In this section, we study both the overlap
and the stability of the two trajectories.

\subsubsection{Overlap of the atomic trajectories}
When using Raman laser beams oriented in the vertical direction,
the atomic trajectories have to be overlapped in the orthogonal
plane (xy) since the Raman laser system remains invariant along
the z direction. In addition, the two atomic clouds must be
resonant simultaneously with the single Raman beams which implies
that vertical velocities must be equal. The key parameters
allowing to fulfill these requirements are the initial positions
and velocities of the two atomic sources. Despite a careful
alignment and power adjustment of the twelve trap laser beams, the
overlap of the two trajectories has to be finely tuned. The
adjustment of the velocities is achieved by modifying the
launching directions~(thanks to the tilt of the experiment) and by
changing independently the values of the launch velocities. The
relative positions at the moment of the Raman pulses were
optimized by adjusting the timing sequence and the positions of
the zeros of the quadratic magnetic fields in the two traps.

\begin{figure}
\includegraphics[width=8cm,angle=0]{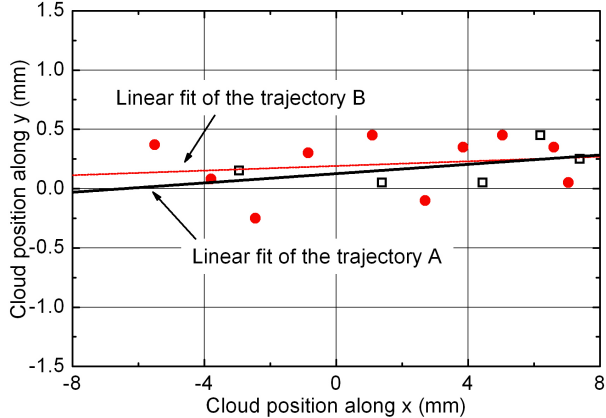}
\caption{(Color online) Trajectories of the atomic clouds A (black
squares) and B (red dots) in the xy plane, in the interaction
area. Two linear fits of experimental data show that the overlap
is better than~0.5~mm, limited by the measurement resolution.}
\label{fig-traj}
\end{figure}

We directly map the two trajectories in the interferometer zone
thanks to the Raman laser beams, with an movable aperture of
diameter~5~mm. By maximizing the transfer efficiency of a Raman
pulse, we deduce the actual position of each atomic cloud in the
xy plane at a given time. Fig.~\ref{fig-traj} shows the two
trajectories which are overlapped to better than~0.5~mm over the
whole interferometer zone. The measurement resolution is limited
by the spatial extension of the atomic clouds (5~mm FWHM at the
apex).

\subsubsection{Stability of the trajectory overlap}
The stability of the overlap is of primary interest for the
stability of the rotation rate
measurements~(paragraph~\ref{Sensitivity}). Ideally, independent
measurements of the  initial velocities and positions of the two
sources in the (xy) plan should be performed. Unfortunately, the
use of the time of flight method and Raman lasers along either the
y or the z axes only gives access to the initial position
stabilities in the z axis and velocities in the y and z
directions. Nevertheless, this measurement gives an evaluation of
the typical fluctuations of position and velocity which are needed
in the analysis of the long term stability.

\begin{figure}
\includegraphics[width=8cm]{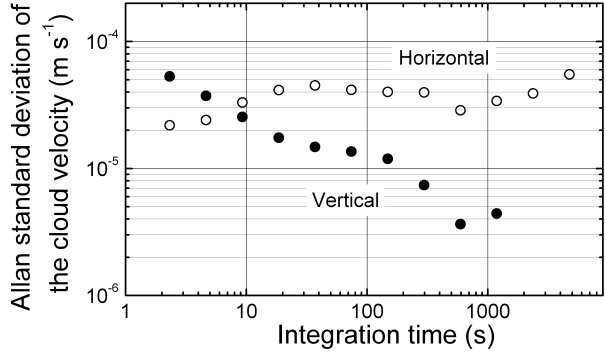}
\caption{Allan standard deviation of the atomic velocity for
source A measured by Raman spectroscopy. The analysis is carried
out from a 12~hour continuous acquisition.}
 \label{stabvit}
\end{figure}

The atomic velocity stability is measured by Raman spectroscopy,
using horizontal or vertical Raman laser beams. For
counter-propagating Raman beams, the resonance condition depends
on the Doppler effect and leads to a measurement of the atomic
velocity. In addition, to discriminate the Doppler effect from
other sources of frequency shift, we use the k-reversal method
described in chapter~\ref{Interf}. Fig.~\ref{stabvit} shows the
Allan standard deviations of velocities along the z and y
directions for source A. Similar behaviors have been recorded for
source B, with stabilities between 2 and 30~$\mu$m~s$^{-1}$ for
time scales from 1 to 5000~s.
In addition, the fluctuations of the initial position along the
vertical direction can be obtained by combining the previous
velocity data with simultaneous time of flight measurements.
Therefore we can estimate the Allan standard deviation of the
initial position, which is 28~$\mu$m at 2 seconds, averaging over
the long times (1000~s) to 15~$\mu$m. Combining these position and
velocity fluctuations independently, we infer a typical overlap
fluctuation of around 20 to 30~$\mu$m in the interferometer zone.
In order to explain the sources of these fluctuations, additional
studies have been performed. They demonstrate that relative power
and polarization fluctuations of the opposing cooling beams can
explain this result. Indeed, we measured a change of the launch
velocity of 36~$\mu$m~s$^{-1}$ for 1\% of polarization fluctuation
and a initial displacement of 50~$\mu$m for 1\% of intensity
fluctuation between the three top and the three bottom cooling
beams~\cite{gravi2008}. In normal operation, polarization
fluctuations are the dominant contribution affecting the stability
of the trajectories.

\subsection{Measurement of the systematic errors and the scale factor}
In order to perform inertial force measurements, we need to know
accurately the scale factor and the systematic errors, which link
the actual rotation and acceleration quantities to the measured
phase shifts $\Delta\Phi_\mathrm{rot}$ and
$\Delta\Phi_\mathrm{acc}$. In order to determine these two
parameters for rotation, we change, in a controlled way, the
rotation rate measured by the gyroscope. In addition, taking
advantage of the single Raman laser beam pair, the interaction
time T can be changed continuously from $T = 0$ to 40~ms. The
systematic error and scale factor measurements were performed for
various interaction times, giving a test of the quadratic scaling
of the rotation phase shift with $T$.

\subsubsection{Dependence on the rotation rate}
The first test of the rotation scale factor consists in checking
the proportionality between the rotation phase shift and the
rotation rate. For this purpose, the orientation $\theta$ between
the sensitive axis of the gyroscope and the East-West direction,
is changed to measure different horizontal projections of the
Earth rotation rate, $\Omega_y =\Omega_h  \sin\theta$. At the
Observatoire de Paris, located at the latitude $\lambda =
48^0\;50'\;08''$, the horizontal rotation rate $\Omega_h$ is
$4.8\times10^{-5}$~rad~s$^{-1}$. The whole device is placed on a
rotation mount, which determines the orientation with a relative
accuracy of 50~$\mu$rad.

\begin{figure}
\includegraphics[width=8cm,angle=0]{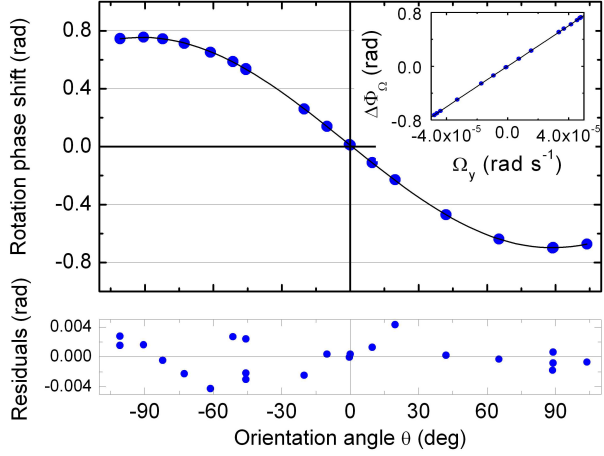}
\caption{(Color online) Measurement of the rotation phase shift as
a function of the orientation $\theta$ of the experimental setup
with respect to the East-West direction. Experimental data (dots)
and their sinusoidal fit (line) are displayed on the top graph
while their residuals are shown at the bottom. The inset exhibits
the same data reported as a function of the rotation rate.}
\label{rot-angle}
\end{figure}

In Fig.~\ref{rot-angle} the rotation phase shift is plotted with
respect to the orientation of the gyroscope. The experimental data
are fitted by a sinusoidal function
${\Delta\Phi~=~\Delta\Phi_{\Omega}^{er}+A^{(0)}\sin(\theta-\theta_0)}$.
The magnitude $A^{(0)}$ is used to calibrate the rotation scale
factor: we find ${A^{(0)}/\left(\Omega_T \sin\lambda\right)=15124
\pm 12\; \mathrm{rad}~\mathrm{rad}^{-1}~\mathrm{s}^{-1}}$.
Moreover this measurement allows to deduce the systematic error of
the rotation signal:
${\Delta\Phi_{\Omega}^{er}~=~28.3~\pm~0.7}$~mrad.

The same data, when plotted as a function of the rotation rate,
are well aligned with a slope equal to the scale factor.
Non-linearities appear as deviations to the straight line and can
be evaluated by fitting with a quadratic term. Their relative
contributions are below $10^{-5}$ in the range of measurement.

\subsubsection{T scaling of the rotation phase shift}
\begin{figure}
\includegraphics[width=7 cm]{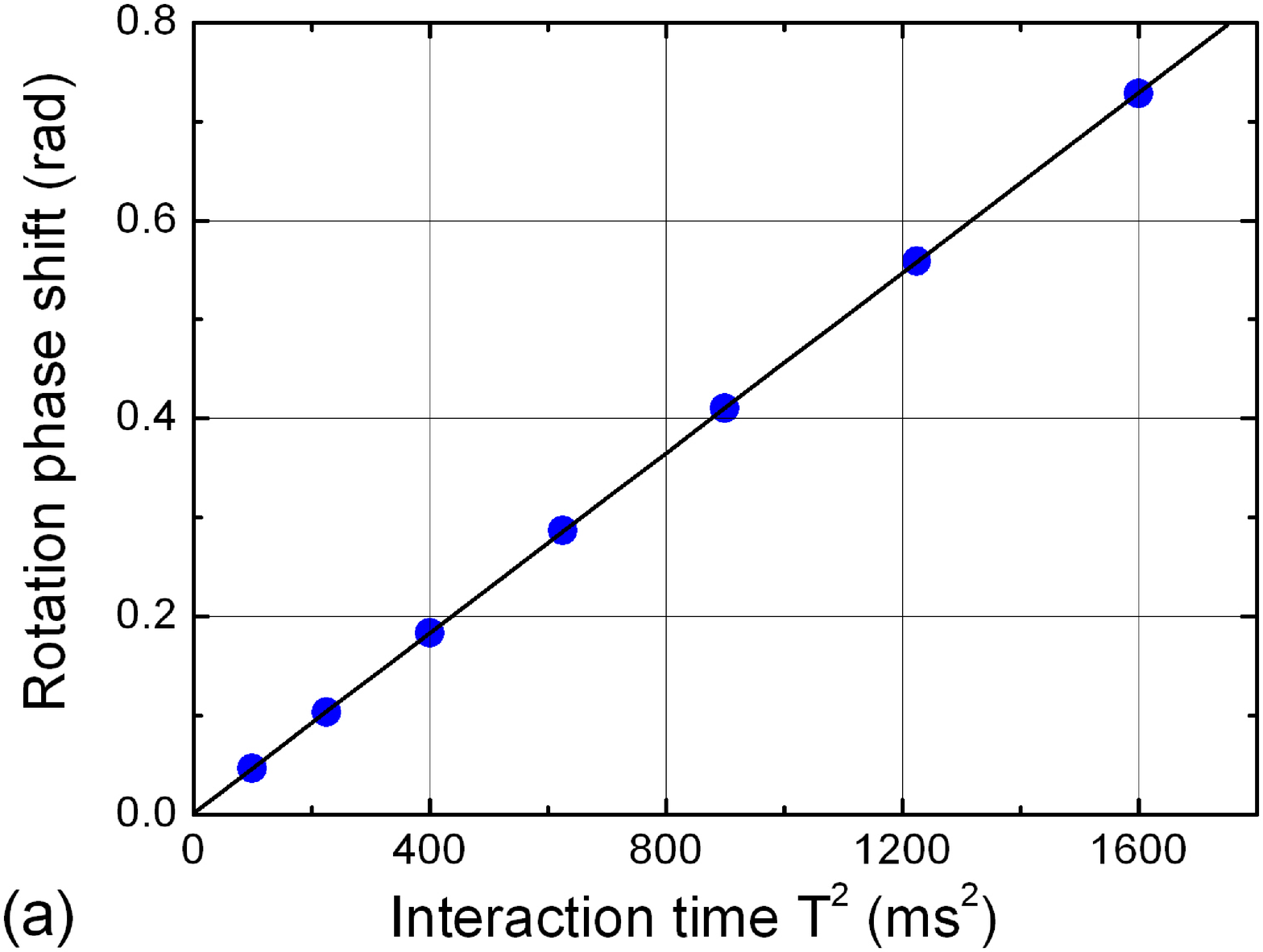}
\includegraphics[width=7 cm]{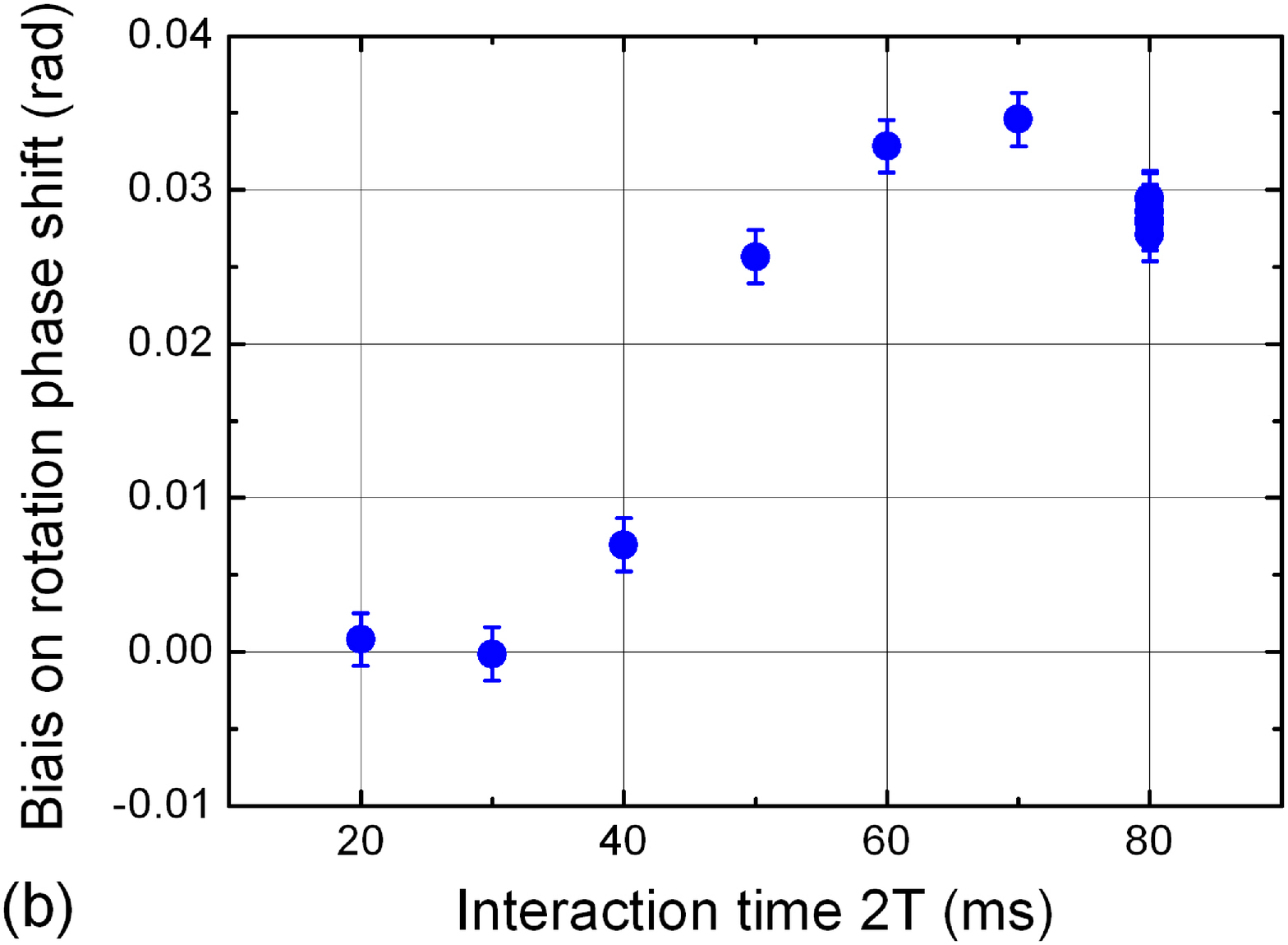}
\caption{(Color online) (a) Evaluation of the rotation phase shift
versus the interaction time $T^2$. In this figure the systematic
error is cancelled by the substraction of the rotation signals
obtained at $\pm~90^{\circ}$. (b) Systematic error on the rotation
signal as a function of the interaction time 2T. The systematic
error is evaluated by calculating the half sum of the rotation
signals measured at~$\pm~90^{\circ}$.} \label{fig-biais-ech-rot-T}
\end{figure}

In order to test the $T^2$ scale factor dependance, the
experimental setup is oriented at ${\theta~\simeq
\pm~90^{\circ}}$, for which the sensitivity of the gyroscope to
the Earth rotation is maximum. We also take advantage of the fact
that, for these two orientations, an error in the orientation
angle $\theta$ has a second order effect on the rotation phase
shift. Therefore we write:
\begin{eqnarray}
\begin{array}{l}
\Delta\Phi_{\Omega}^{+90^{\circ}}(T) = \Phi_{rot}^{er}(T)-2k_{\mathrm{eff}}V_x\Omega_y T^2\\[0.5cm]
\Delta\Phi_{\Omega}^{-90^{\circ}}(T) =
\Phi_{rot}^{er}(T)+2k_{\mathrm{eff}}V_x\Omega_y T^2
\end{array}
\end{eqnarray}

where $\Phi_{rot}^{er}(T)$ is the systematic error on the rotation
signal for a given interaction time~$T$. The half difference of
the signal measured at $\theta = +90^{\circ}$ and
$\theta~=~-90^{\circ}$~(Fig.~\ref{fig-biais-ech-rot-T}(a)) gives a
test of the scaling of the rotation phase shift as $T^2$. It shows
an excellent agreement with the expected behavior. In addition,
the systematic error which affects the rotation signal is deduced
from the half sum of the signals for the two orientations. The
evolution of the error phase shift is displayed as a function of
the interaction time in Fig.~\ref{fig-biais-ech-rot-T}(b).
Negligible for small interaction times, it increases up
to~34~mrad. For the usual interaction time of~$2T~=$~80~ms, this
systematic error reaches~28.3~mrad. The study of the different
sources of systematic errors is performed in the following
section.

\subsection{Investigation of the sources of systematic errors}\label{Systematic errors estimation}
\subsubsection{Quadratic Zeeman effect}
The first expected source of systematic error in the
interferometer phase shift comes from the Zeeman shift induced by
the magnetic field. In order to limit its impact, the
interferometer is realized between the two~${m_F=0}$ Zeeman
sub-levels, whose energy difference evolves quadratically with the
magnetic field as: $\Delta_{mag} = K^{(2)}B^2$, with
$K^{(2)}~=~427.45$~Hz~G$^{-2}$ for Cesium atoms. In addition,
thanks to the symmetric features of the interferometer, the phase
shift is not sensitive to a constant frequency shift. However, a
magnetic field gradient breaks the symmetry and gives rise to a
phase shift. Assuming a linear gradient $\delta_{bx}$ along the
atomic trajectories the phase shift induced is:
\begin{equation}
\Delta \Phi_{mag} = -4 \pi K^{(2)} B_0 T^2 V_{x} \delta_{bx}
\end{equation}

\begin{figure}
\includegraphics[width=8cm,angle=0]{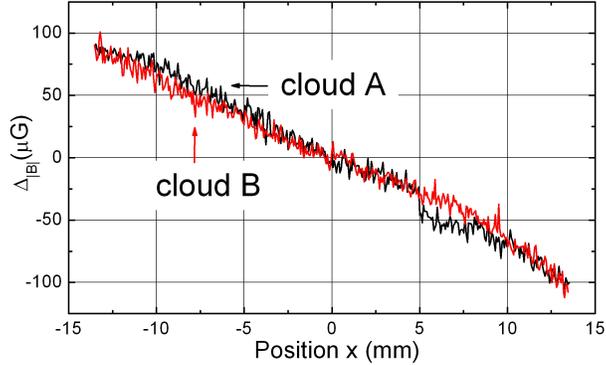}
\caption{(Color online) Measurement of the magnetic field as a
function of the position of the atoms along their trajectory in
the interaction zone.} \label{B-field}
\end{figure}

The MOTs and the interaction region are set into three independent
magnetic shields and the whole experiment is inside a second layer
of shield so as to limit the influence of the ambient magnetic
field. The magnitude of the residual magnetic field has been
measured by selecting the atoms in the~${m_F=2}$ sub-level and
driving magnetic transitions thanks to a microwave antenna. By
measuring the frequency of the transition between the
${|6S_{1/2},F=3,m_F=2\rangle}$ and
${|6S_{1/2},F'=4,m_{F'}=2\rangle}$ states along the trajectories,
we mapped the value of the magnetic field in the interaction zone.
The results of these measurements are displayed in
Fig.~\ref{B-field} for the two atomic clouds. We measured a
gradient of $\delta_{bx}=7$~mG~m$^{-1}$ giving a small phase shift
of $0.6$~mrad on the rotation signal. This Zeeman phase shift does
not depend on the laser wave vectors, therefore it cancels out
with the k-reversal technique and so disappears on the total
interferometer phase shift.

\subsubsection{AC Stark shift}
\label{AC stark shift}

We study the effect of a frequency shift induced by the AC Stark
shift~$\Delta_{AC}$ on the interferometer phase
shift~\cite{weiss1994}. This frequency shift between the two
hyperfine levels can be cancelled by setting a proper intensity
ratio between the two Raman beams. However, if the ratio is
flawed, a phase shift~$\Delta\Phi_{AC}$ is induced on the
interferometer. Assuming the AC Stark shift to be constant during
each laser pulse gives:
\begin{equation}
\label{ls1} \Delta\Phi_{AC} = \left(
\frac{\Delta_{AC}^{(3)}}{\Omega_\mathrm{eff}^{(3)}}-\frac{\Delta_{AC}^{(1)}}{\Omega_\mathrm{eff}^{(1)}}\right)
\end{equation}

where $\Delta_{AC}^{(1,3)}$ are the frequency shifts of the Raman
transition at the time of the first and the third pulses, and
$\Omega_\mathrm{eff}^{(1,3)}$ are the respective effective Rabi
frequencies~\cite{moler1992}. In our experimental setup, the two
Raman lasers are provided by a single optical fiber and are
retro-reflected, ensuring the stability of the intensity ratio
between the two lasers throughout the Raman beam.

Equation~(\ref{ls1}) shows that if the atomic trajectories are
perfectly overlapped and symmetric with respect to the center of
the Raman beam, the phase shift~$\Delta\Phi_{AC}$ remains equal to
zero as the contributions of the first and last pulses are
identical. However, a parasitic phase shift appears on the
acceleration signal if the gaussian profile of the Raman beam is
not centered with respect to the three pulse sequence of the
interferometer. A similar effect would be induced by time
fluctuations of the laser power at the output of the fiber.
Moreover, this phase shift also affects the rotation signal if the
two trajectories are not perfectly superimposed.

The sensitivity of the rotation signal to the AC Stark shift was
characterized by changing the intensity ratio between the two
Raman lasers so as to induce a frequency shift
of~$\Delta_{AC}=10$~kHz. A sensitivity of 5~mrad~kHz$^{-1}$ was
measured. This AC Stark shift is however independent of the
effective wave vector $\textbf{k}_\mathrm{eff}$ so it can be
rejected by the k-reversal method, considering that the
fluctuation of~$\Delta\Phi_{AC}$ is slow compared with the
repetition rate of the measurement. Using this technique, we
measure a residual error due to the AC Stark shift at the level
of~0.1~mrad~kHz$^{-1}$ corresponding to a sensitivity to this
effect reduced by a factor of 50. When running the interferometer,
we directly optimize the AC Stark shift on the atomic signal with
an accuracy better than 500~Hz, leading to a residual systematic
error below 0.05~mrad which is negligible.

\subsubsection{Two-photon light shift}
As explained above the retro-reflected Raman beams couple the
ground state ${|6S_{1/2},F=3,\textbf{p}\rangle}$ with the two
states ${|6S_{1/2},F=4,\mathbf{p}\pm\hbar
\mathbf{k}_{\mathrm{eff}}\rangle}$. Since these two possible Raman
transitions are Doppler shifted, we can choose only one state by
adjusting the Raman detuning. However, the non-resonant coupling
induces a two-photon light shift (TPLS) on the selected Raman
transition which results in an atomic phase shift
($\Delta\Phi_{TPLS}$)~\cite{gauguet2008}:
\begin{equation}
\Delta\Phi_{TPLS} =
\frac{\Omega_\mathrm{eff}^{(1)}}{4\delta_D^{(1)}}-\frac{\Omega_\mathrm{eff}^{(3)}}{4\delta_D^{(3)}}
\end{equation}

where $\Omega_\mathrm{eff}^{(i)}$ is the effective Rabi frequency
and $\delta_D^{(i)}$ the Doppler shift for the $i^{th}$ pulse.
This phase shift depends on $\mathbf{k}_{\mathrm{eff}}$ and cannot
be cancelled out with the k-reversal method. It is shown that this
shift is similar for the two interferometers and induces an error
on the acceleration signal of only~12~mrad while remaining below
0.3~mrad on the rotation signal.

\subsubsection{Wave-front distortion}
\label{Wave-front distortion} The atomic phase shift depends on
the effective laser phases~$\phi_i$ imprinted on the atomic wave
at the moments of the three pulses. Because of a non-uniform laser
wave-front, the phase shift measured by each interferometer
depends on the position of the atomic cloud in the Raman
wave-front~(${x_i, y_i}$) at the i$^{th}$ pulse and can be written
as:
\begin{equation}
\Delta\Phi_\mathrm{wf} =
\phi_1(x_1,y_1)-2\phi_2(x_2,y_2)+\phi_3(x_3,y_3)
\end{equation}
The spatial variations of the laser phase along the wave-front
induce a phase shift on the interferometric measurements. In our
setup, the use of a retro-reflected configuration allows to reduce
the number of optical elements and hence to decrease the
aberrations between the two opposite Raman beams. Therefore, the
wave-front defects are mainly induced by the Raman window, the
quarter wave plate and the retro-reflection mirror, which affect
only the reflected beam.

When the trajectories of the two atomic clouds are perfectly
overlapped, the wave-front defects are identical for both
interferometers, which is equivalent to a constant acceleration.
\begin{figure}
\includegraphics[width=8cm]{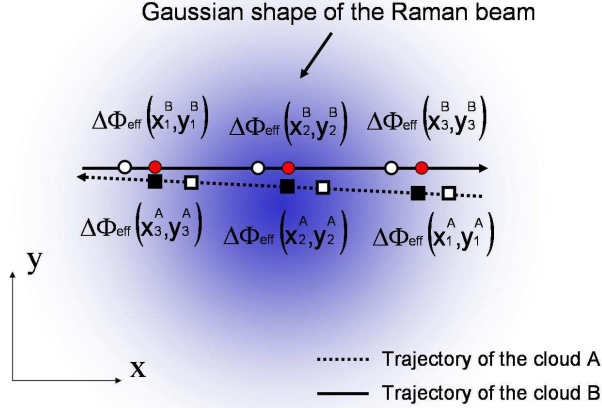}
\caption{(Color online) Wave-front defects induce a phase shift
related to the laser phases seen by the atoms at the moments of
the three laser pulses (black squares and red dots). By changing
the moments when the pulses occur we simulate a relative
displacement of the atomic clouds with respect to the Raman beam
(white squares and white dots).} \label{aberration}
\end{figure}
If the paths are not perfectly overlapped, the phase shifts due to
these wave-front defects are not identical and appear as an error
on the rotation signal when subtracting the phase shifts of the
two interferometers~\cite{fils2005}. This is illustrated in
Fig.~\ref{aberration}, which represents the positions of the
atomic clouds at the times of the three pulses $\pi/2-\pi-\pi/2$.
The error on the rotation signal results from the half difference
of the two wave-front distortion phase shifts: $ \Delta
\Phi_\mathrm{wf} =
\frac{\Delta\Phi_\mathrm{wf}^{A}-\Delta\Phi_\mathrm{wf}^{B}}{2} $.

\begin{figure}
\includegraphics[width=8cm]{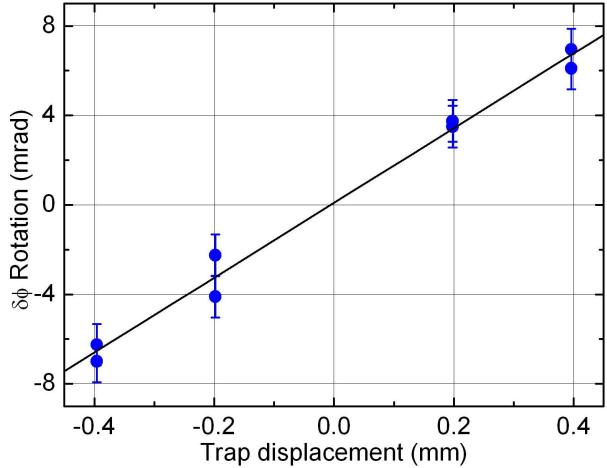}
\caption{(Color online) Rotation phase shift measured as a
function of the atomic cloud positions along the x direction. The
displacements of the atomic position compared to the laser beam
are applied by changing the timing of the laser pulse sequence.}
\label{fig-rotation-tim}
\end{figure}

A first estimation of this effect was performed by measuring the
wave-front distortion induced by a single Raman window with a Zygo
wave-front analyzer~\cite{Fils2002}. From these measurements, we
deduce a wave-front phase shift of~20~mrad on the rotation signal
for an interaction time of $2T=80$~ms, which is consistent with
the error measured on the rotation signal. A second method was
implemented to estimate this effect by moving the
positions~${(x_i, y_i)^{A,B}}$ of the atomic clouds in the Raman
beam at the moments of the pulses. This can be performed by
changing the delay between the launch and the first Raman pulse,
keeping the duration of the interferometer constant~($2T =
80$~ms). Consequently, the positions of the two atomic clouds are
shifted in opposite directions, as shown on~Fig.~\ref{aberration}.
In Fig.~\ref{fig-rotation-tim}, the rotation phase is displayed as
a function of the delay compared to the usual laser sequence, and
translated into trajectory shift.
For small deviations, the phase shift due to wave-front defects
can be linearized. Therefore we infer the sensitivity of the
rotation measurement to a relative displacement along the x
direction between the two sources by performing a linear fit of
the data. The sensitivity obtained is:
\begin{equation}
\frac{\Delta\Phi_\mathrm{wf}}{\delta x} = 17
\;\mu\mathrm{rad}~\mu\mathrm{m}^{-1}\label{scalefact-aberation}
\end{equation}

\subsubsection{Conclusion on systematic errors}

We studied the relative contributions of different sources of
parasitic phase shifts. The dominant contribution to the
systematic effects has been identified as coming from the
wave-front distortions. Indeed, an independent evaluation of its
contribution~($\sim$~20~mrad) is in agreement with the actual
parasitic phase shift~(28.3~mrad) for
$2T=~80$~ms~(Fig.~\ref{fig-biais-ech-rot-T}(b)). This effect
becomes significant when the interrogation time exceeds~30~ms. We
attribute this to the growth of wave-front distortions on the
edges of the windows. Using a larger window or the centers of
three separated windows for the three pulses would limit its
impact.

\section{Sensitivity}\label{Sensitivity}

The bias stability of the acceleration and rotation signals was
studied by orienting the area of the interferometer in the
East-West direction so that the rotation rate measured is zero.
Consequently, it is possible to measure a phase shift by setting
the interferometers on the side of the fringe for both
interferometers simultaneously. Then we calculate the phase shift
from the measured probabilities $P^{A,B}$, the contrast $C^{A,B}$
and the offsets~$M^{A,B}$.

Since the interferometers operate at the side of a fringe, the
fluctuations of the contrast parameters $C^{A,B}$ do not impact
significantly the measured probabilities. Consequently, the
contrast values are determined once at the beginning of the
measurement, by fitting the fringe pattern with a sinusoidal
function. To eliminate offset fluctuations, the experimental
sequence alternates measurements on both sides of the central
fringe. The half-difference between two successive measurements
yields the atomic phase shift rejecting the offset fluctuations.
Moreover, the sign of the effective wave vector
$\mathbf{k}_\mathrm{eff}$ is reversed between two successive
steps.

\begin{figure}
\includegraphics[width=8cm]{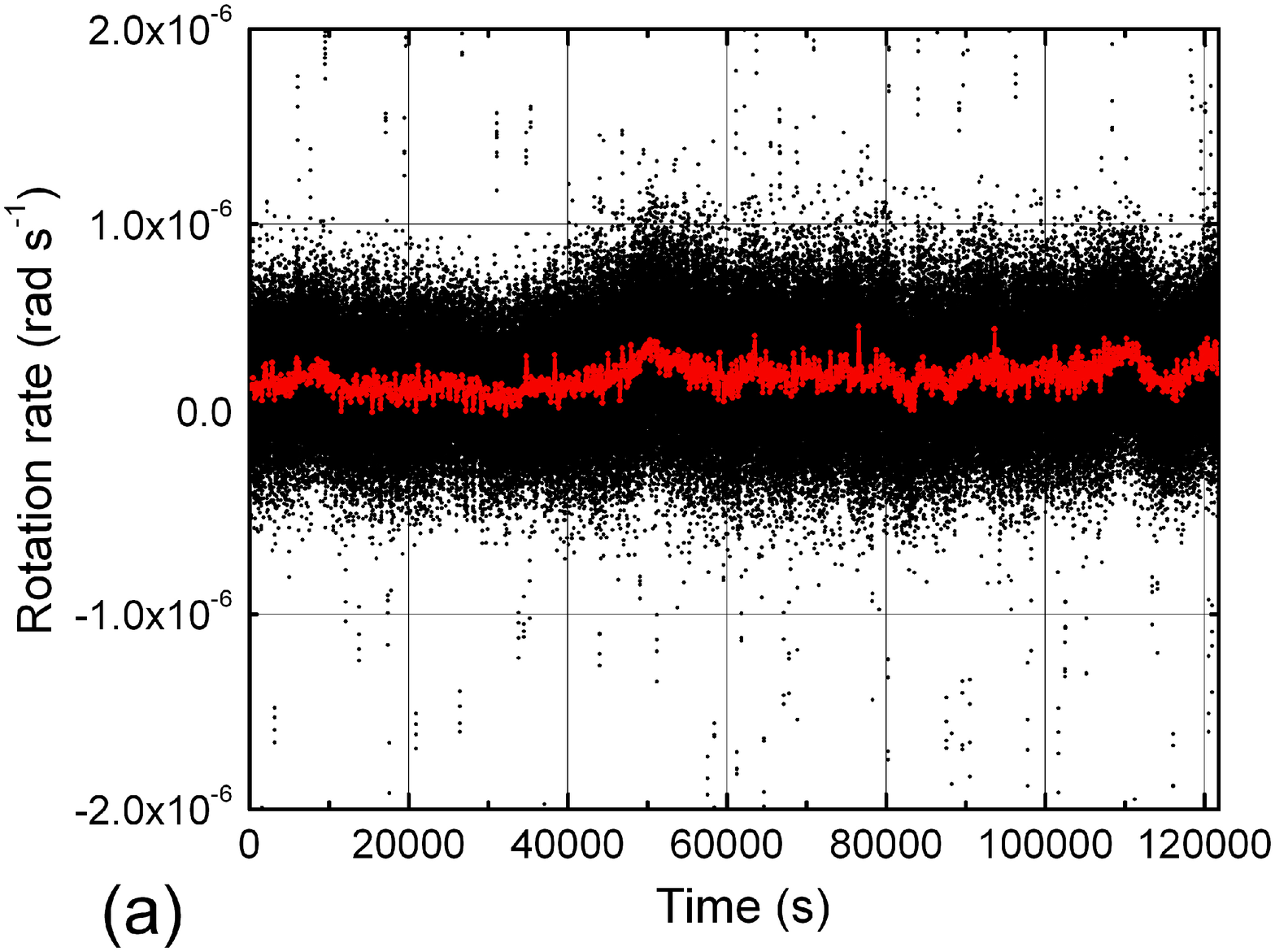}
\includegraphics[width=8cm]{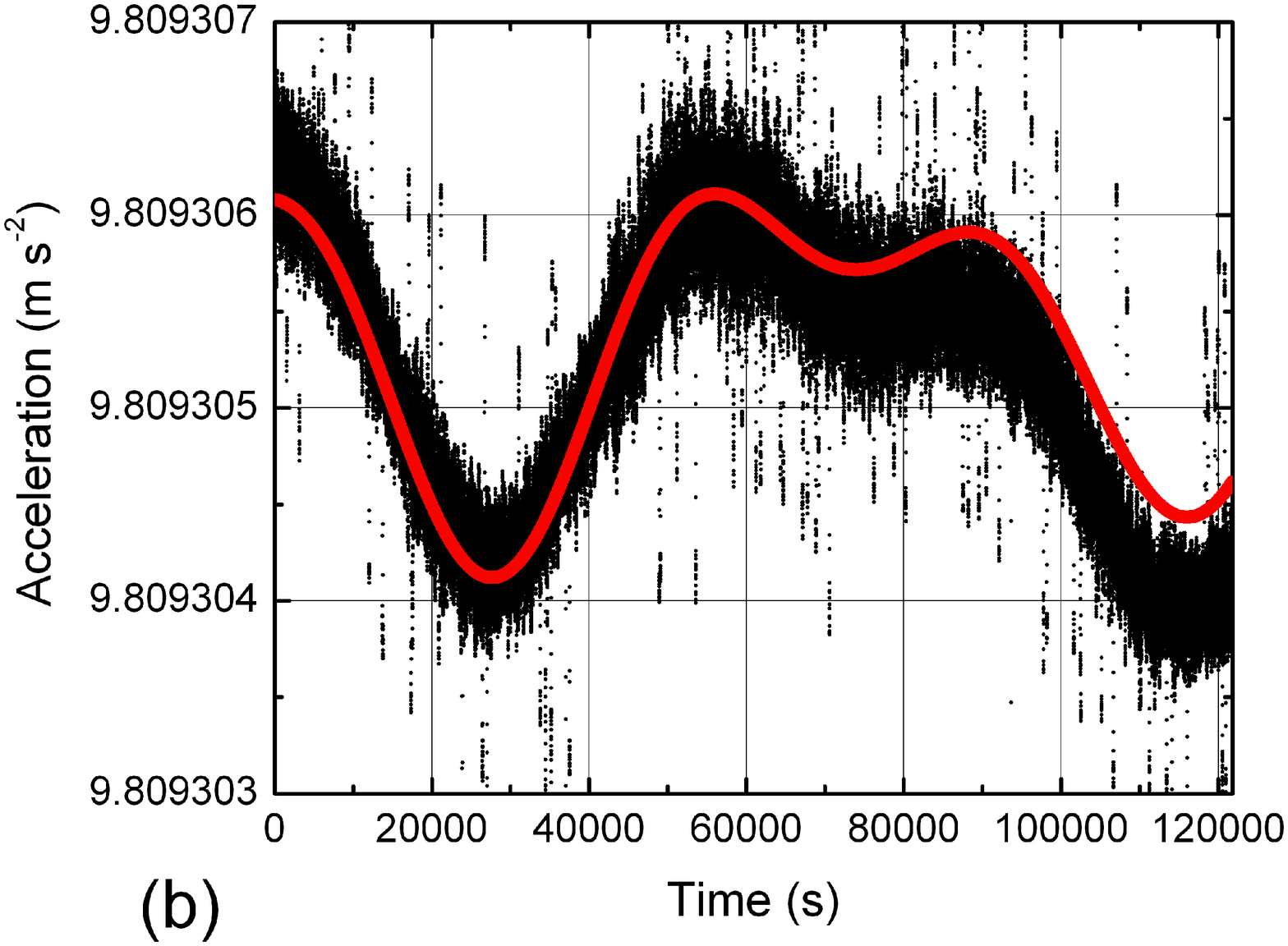}
\caption{(Color online) Rotation (a) and acceleration (b)
measurements as a function of the time, obtained for an
interaction time of ${2T=80}$~ms on a 12~hours continuous
acquisition. The rotation data~(a) are presented shot-to-shot
(black dots) and averaged over 100~s (red line). The  shot-to-shot
acceleration measurement (b) is compared to expected variations of
the gravity (line) due to tidal effects.}
\label{signaux_temporels}
\end{figure}

Fig.~\ref{signaux_temporels} shows the two inertial signals as a
function of time, for an interaction time of $2T=80$~ms and a
repetition rate of 1.72~Hz. The large oscillations which appear on
the acceleration signal are due to tidal effects and are removed
on the rotation signal illustrating a key feature of our geometry.
A more quantitative study of the separation between the
acceleration and rotation signal demonstrated a 76~dB rejection
rate, the details of this study are presented in
appendix~\ref{separationaccrot}. The efficiency of this rejection
is due to the fact that the wave packets of the two sources
interact with the same equi-phase plane of the Raman beams at
exactly the same time.

\subsection{Acceleration measurements}
The acceleration signal is deduced from the sum of the two
interferometer phase shifts. The short term sensitivity obtained
on these measurements is
${5.5\times10^{-7}}$~m~s$^{-2}$~Hz$^{-1/2}$. This sensitivity is
mainly limited by the residual vibrations of the platform, as for
the atomic gravimeter~\cite{legouet2008}. The standard
deviation~(Fig.~\ref{fig-stab}(a)) of the acceleration signal
shows an improvement of the sensitivity proportional to
$\tau^{-1/2}$ as expected. By integrating our measurement over
5000~s, we reach a sensitivity of $10^{-8}$~m~s$^{-2}$, which is
close to our atomic gravimeter characteristics presented
in~\cite{legouet2008}. The difference between the sensitivity of
the two apparatuses is explained by our slightly lower repetition
rate and interrogation time. In order to reach this sensitivity,
the noise contribution from vibrations is filtered out thanks to a
passive isolation platform (nanoK 350BM-1). The residual noise is
further reduced by a correlated measurement performed with a low
noise seismometer~(Guralp~T40)~\cite{legouet2008}. Additionally,
variations of the gravity~$g$ due to tidal effects are computed
from a model provided by tide parameters extracted
from~\cite{tidal} and subtracted from the signal in order to infer
the long term stability of the sensor.

\subsection{Noise on the Rotation signal}
\begin{figure}
\includegraphics[width=8cm,angle=0]{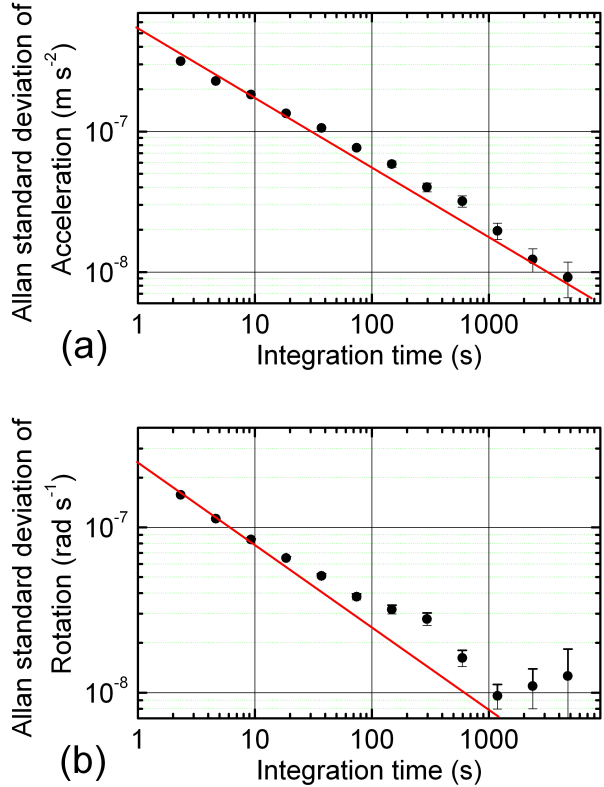}
\caption{(Color online) Allan standard deviation calculated for
acceleration~(a) and rotation~(b) measurements (dots) from a
36~hour continuous acquisition. The lines show the expected
improvement of the sensitivity proportional
to~$\tau^{-1/2}$}.\label{fig-stab}
\end{figure}

The rotation phase shift is extracted from the difference between
the signals of the two interferometers. The Allan standard
deviation is plotted in Fig.~\ref{fig-stab}(b). The short term
sensitivity of the rotation signal is 2.4$\times 10^{-7}\;
\mathrm{rad~s}^{-1}~\mathrm{Hz}^{-1/2}$. The Allan standard
deviation decreases with integration time as $\tau^{-1/2}$, down
to 1000 seconds, reaching a sensitivity of
$10^{-8}\;\mathrm{rad~s}^{-1}$.

The Allan standard deviation of the rotation signal at one second
is limited by the quantum projection noise evaluated in section
\ref{detection}. In order to confirm this point, we perform
measurements for different numbers of atoms by changing the
loading time of the two MOTs. Assuming that the detection noise is
independent for the two interferometers A and B, its impact on the
standard deviation $\sigma_{\Phi}$ of the rotation phase shift
yields:
\begin{eqnarray}
\label{bruit-phase-detect} \sigma_{\Phi}^2  = \frac{1}{4C^2}
\Bigl( 2\alpha+ \frac{1}{4}(\frac{1}{N^{A}}+\frac{1}{N^B})+
\gamma(\frac{1}{{N^A}^2}+ \frac{1}{{N^B}^2}) \Bigr)
\end{eqnarray}

The contrasts of the two interference fringes are identical for
the two interferometers, and denoted by $C$. The coefficients
$\alpha$ and $\gamma$ are related to the detection features and
were determined in section \ref{detection}. In order to
characterize the contribution of the detection noise, it is
convenient to plot the rotation phase noise versus the number of
atoms. For a given loading time, the number of trapped atoms is
different for the two atomic sources A and B, so we define a
reduced atom number ${N = \frac{N_A N_B}{N_A + N_B}}$. Thus
equation~(\ref{bruit-phase-detect}) becomes:
\begin{eqnarray}
\label{bruit-phase-detect-ap} \sigma_{\Phi}^2 \simeq
\frac{1}{4C^2}\bigl(2\alpha+\frac{1}{4N}+\frac{\gamma}{N^2}\bigr)
\end{eqnarray}

\begin{figure}
\includegraphics[width=8cm,angle=0]{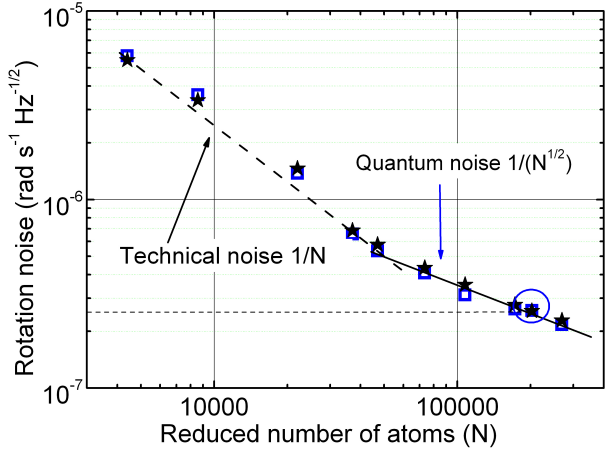}
\caption{(Color online) Rotation noise at 1 second measured on the
interferometer (blue squares) and estimated taking into account
the parameters of the detection system (stars) as a function of
reduced atom number. The circle shows the usual parameters of the
experiment corresponding to a reduced number of~$2 \times 10^{5}$
atoms per shot.} \label{bruit-rot}
\end{figure}

In Fig.~\ref{bruit-rot} the rotation noise (blue squares) is
displayed versus reduced atom number. The data correspond to the
Allan standard deviation at 1 second of the rotation phase shift
calculated from series of 10~minutes of measurements. The black
stars correspond to the noise estimated from equation
(\ref{bruit-phase-detect-ap}) in which the coefficients
$\alpha,\gamma$ were measured independently of the interferometric
measurement (section \ref{detection}). We note an excellent
agreement between rotation noise measured with the two
interferometers and the detection noise evaluated independently.
The number of atoms in the usual conditions is indicated by the
circle in Fig.~\ref{bruit-rot}, which corresponds to a rotation
noise of 2.4$\times
10^{-7}\;\mathrm{rad~s}^{-1}~\mathrm{Hz}^{-1/2}$, limited by the
quantum projection noise. An improvement by a factor of 45 on the
atom number ($\sim 10^{7}$ atoms), would allow to reach the
detection noise plateau independent of the atom number, estimated
from the parameter $\gamma$ to be~$4\times 10^{-8}\;
\mathrm{rad~s}^{-1}$~Hz$^{-1/2}$. Such a number of atoms is now
obtained during the usual loading time of our interferometer by
using a 2D MOT~\cite{Dieckmann1998}.

\subsection{Long term stability}
The long term sensitivity achieves a plateau at
$10^{-8}$~rad~s$^{-1}$ for time scales longer than 1000~s. We have
carried out a systematic study of all possible sources of drift,
which can limit the sensitivity for longer measurement times.
First, we have verified that the orientation of the sensitive
rotation axis is stable in space. Second, we have quantified the
effect of a possible drift from every systematic error source.
Only the fluctuations of trajectories, coupled to the wave-front
distortions of the Raman laser, can explain the observed drift.
The typical position fluctuations~(20~$\mu$m) and drift
sensitivity of $10^{-9}$~rad~s$^{-1}$~$\mu$m$^{-1}$ gives a
typical limit of the order of $2.10^{-8}$~rad~s$^{-1}$, in
agreement with the observed value.

This main limitation can be reduced drastically by combining an
improvement of the wave-front of the Raman laser and a reduction
of the position fluctuations. Placing the retro-reflection system
($\lambda/4 \;+$ mirror) inside the vacuum chamber removes
aberrations induced by the window, which represents the highest
contribution. The stabilities of the trajectories can be improved
by the use of other kinds of optical fibers (having better
stability of the polarization) or an active stabilization of
polarization and intensity of each cooling beam.

\section{Conclusion}
We carried out the characterization of a cold atom gyroscope in
terms of sensitivity, systematic errors and scale factor. A study
of the scale factor demonstrated excellent linearity and
stability, limited by the resolution due to the drift of the
systematic effects. This first study of the limits of a gyroscope
using cold atoms has already demonstrated a sensitivity at the
level of the best commercial optical gyroscopes (fiber and ring
laser gyroscopes). In other work, a 400~times better short term
performance was demonstrated using an atomic beam interferometer
\cite{Gustavson2000}. However, considering both at short and long
term sensitivity, our gyroscope is~3~times less sensitive than the
best atomic one~\cite{Kasevich2006}. Moreover, this work has
clearly identified the limits to the sensitivity, pointing the way
to further improvements.

The short term sensitivity was dominated by the quantum projection
noise thanks to the use of a double-interferometer, which
perfectly cancels the phase shift due to parasitic vibrations. The
main contribution to the drift is related to the fluctuations of
the atomic trajectories. When coupled to the Raman wave-front
distortions, these fluctuations also limit the long term stability
of the rotation measurements.  More generally, similar effects
from wave-front distortions should appear in the other kinds of
dual cold atom interferometers, based on molasses techniques, such
as the gravity gradiometers~\cite{mcguirk2002,lamporesi2008}, or
on in tests of the universality of free fall by comparing the
accelerations of two clouds of different
species~\cite{nyman2006,ertmer2009}.

Finally, these limits are not fundamental and can be improved by
at least one order of magnitude through various improvements. The
parasitic shifts due to wave-front distortions can be reduced by
improving the quality of the optics and the stability of the
launch velocities. Furthermore, their impact in terms of rotation
rate can be reduced by modifying the geometry. Indeed, atoms can
be launched in straighter trajectories with a higher longitudinal
velocity as in Ref.~\cite{rasel2009} or by using the four pulse
configuration previously demonstrated in Ref.~\cite{canuel2006},
with a longer interaction time. In both cases, the area of the
interferometer is significantly increased while keeping the phase
shift due to wave-front distortions almost constant. Long term
performance should then be improved to reach an expected level
below 10$^{-10}$~rad~s$^{-1}$, as achieved with giant ring laser
gyroscopes~\cite{giantRinglaser}, opening the way to new fields of
application for atomic gyroscopes.

\begin{acknowledgments}
We would like to thank Ch.~Bord\'{e}, P.~Bouyer, A. Clairon,
N.~Dimarcq, J.~Fils, D.~Holleville, P.~Petit, F.~Yver-Leduc, who
contributed to build the setup in the early stage of the
experiment. We also thank F.~Pereira Dos Santos for fruitful
discussions and P.~Tuckey for careful reading. We thank the
Institut Francilien pour la Recherche sur les Atomes Froids
(IFRAF) and the European Union (FINAQS STREP/NEST project contract
no 012986 and EuroQUASAR/IQS project) for financial support. T.L.
thanks the DGA for supporting his work. W.C. thanks the IFRAF for
supporting his work.
\end{acknowledgments}

\appendix

\section{Separation of acceleration and rotation phase shifts}
\label{separationaccrot} As our apparatus measures simultaneously
two independent effects (acceleration and rotation), it is crucial
to quantify how these two phase shifts are discriminated by the
dual interferometer technique. Since the experiment enables the
measurement of the six axes of inertia~\cite{canuel2006}, this
characterization was carried out in the horizontal Raman
configuration which enables us to easily vary the acceleration
phase shift over a large range and in a controlled way. The best
performances, using this configuration, have been presented in
Ref.~\cite{RFM2007} and are summarized in the following section.

\subsection{Horizontal Raman beam configuration}

In the horizontal configuration, the retro-reflected Raman beam is
orientated in the y direction so as to realize the two
interferometers in the (xy) plane at the apex of the trajectories.
Therefore the interferometric phase shift is sensitive to the
horizontal acceleration $a_y$ and the vertical rotation rate
$\Omega_z$. When the Raman beams are perfectly horizontal, the
absolute value of the acceleration measured by the interferometer
is close to zero while the rotation measurement records the
vertical component of the Earth rotation rate
$\Omega^{E}_{z}=~5.49 \times 10^{-5}$~rad~s$^{-1}$.
\begin{figure}
\includegraphics[width=8cm,angle=0]{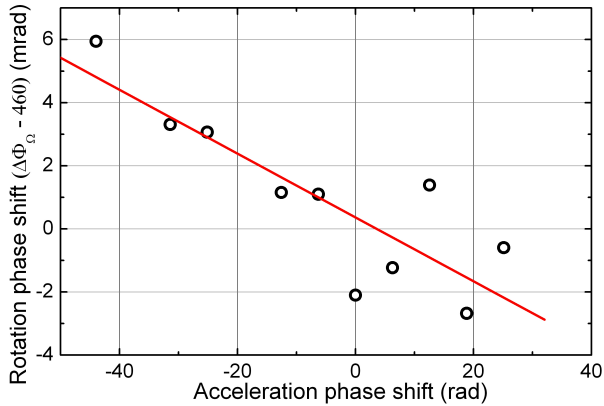}
\caption{(Color online) Measurement of the rotation phase shift as
a function of the acceleration induced by varying the inclination
of the interferometer plane. The measurements are performed in the
horizontal Raman configuration with interferometers of $2T=60$~ms
of total interaction time. Each point corresponds to a measurement
achieved for a given angle $\varepsilon$.}
\label{rejection_acceleration}
\end{figure}

This configuration allows inertial measurements with a total
interaction time up to $2T=60$~ms. Data are acquired in a similar
way as in the vertical configuration, alternating measurements on
the two sides of the central fringe, and for two opposite
effective wave vectors. The short term sensitivity to rotation is
$5.5 \times 10^{-7}$~rad~s$^{-1}~Hz^{-1/2}$. The results are
similar to those obtained on the vertical axis, taking in account
the shorter interaction time T and the reduced contrast (20\%
instead of 30\%).

\subsection{Test of the separation}

The horizontal configuration is well adapted to the measurement of
the rejection of the acceleration phase shift on the rotation
measurement. Indeed, as the interferometer is realized in the
horizontal plane, it is possible to induce a large controlled
change of the acceleration by tilting the device by an angle of
$\varepsilon$ with respect to the horizontal direction. The
interferometer then measures a residual component of the
gravitation $g$ given by:

\begin{equation}
\Delta \Phi_a = k_\mathrm{eff} g \sin{\varepsilon}\ T^2
\end{equation}

By tilting the interferometer plane over a range of 0.5~mrad, we
change the acceleration phase shift from ${\Delta \Phi_a=
-45}$~rad~to~25~rad. Fig.~\ref{rejection_acceleration} displays
the rotation phase shift as a function of the acceleration induced
on the interferometer. The measurements exhibit a very small slope
of~${1.5 \times 10^{-4}}$. Thus the effect of the acceleration on
the rotation signal is cancelled at a level better than 76~dB.

This measurement demonstrates the efficiency of a dual
interferometer gyroscope for applications in the presence of a
relatively high level of acceleration noise.



\begin{thebibliography}{30}

\bibitem{will} C.M. Will. The confrontation between general relativity and experiment. \emph{Living Reviews in Relativity}, \textbf{9}(3) (2006).

\bibitem{stedmann} H. Igel, A. Cauchard, J. Wassermann, A. Flaws, U. Schreiber, A. Velikoseltsev, and N.P. Dinh, Geophys. J. Int. \textbf{168}, 182 (2007).

\bibitem{navigation} A. Lawrence, \emph{Modern Inertial Technology}
(Springer, New York, 1998).

\bibitem{clauser} J.F. Clauser,  Physica B {\bf 151}, 262 (1988).

\bibitem{Riehle91}F. Riehle, Th. Kister, A. Witte, J. Helmcke and Ch.J. Bord\'{e},  Phys. Rev. Lett. \textbf{67}, 177 (1991).

\bibitem{Chu91} M. Kasevich and S. Chu, Phys. Rev. Lett. \textbf{67}, 181 (1991).

\bibitem{Gustavson97} T. L. Gustavson, P. Bouyer, M. A. Kasevich, Phys. Rev. Lett. \textbf{78}, 2046 (1997).

\bibitem{Pritchard97} A. Lenef, T. D. Hammond, E. T. Smith, M. S. Chapman, R. A. Rubenstein, and D. E. Pritchard, Phys. Rev. Lett. \textbf{78}, 760 (1997).

\bibitem{Gustavson2000} T. L. Gustavson, A. Landragin, M. A. Kasevich, Class. Quantum Grav. \textbf{17}, 1-14 (2000).

\bibitem{Kasevich2006} D. S. Durfee,Y. K. Shaham, and M. A. Kasevich, Phys. Rev. Lett. \textbf{97}, 240801 (2006).

\bibitem{canuel2006} B. Canuel, F. Leduc, D. Holleville, A. Gauguet, J. Fils, A. Virdis, A. Clairon, N. Dimarcq, Ch.J. Bord\'e, A. Landragin, and P. Bouyer, Phys. Rev. Lett. \textbf{97}, 010402 (2006).

\bibitem{wu2007} S. Wu, E. Su, and M. Prentiss, Phys. Rev. Lett. \textbf{99}, 173201 (2007).

\bibitem{wang2007} P. Wang, R.B. Li, H. Yan, J. Wang, and M.S. Zhan, Chin. Phys. Lett. \textbf{24}, 27 (2007).

\bibitem{rasel2009} T. M\"{u}ller, M. Gilowski, M. Zaiser, T. Wendrich, W. Ertmer, and E.M. Rasel, Eur. Phys. J. D \textbf{53}, 273 (2009).

\bibitem{borde1989} Ch. J. Bord{\'e}, Phys. Lett. A \textbf{140}, 10-12 (1989).

\bibitem{antoine} Ch. Antoine, and Ch. J. Bord\'{e}, J. Opt. B:
Quantum Semiclass. Opt. \textbf{5}, S199-S207 (2003)

\bibitem{baillard06} X. Baillard, A. Gauguet, S. Bize, P. Lemonde, P. Laurent, A. Clairon, and P. Rosenbusch, Opt. Commun. \textbf{266}, 609 (2006).

\bibitem{muwave} F. Yver-Leduc, P. Cheinet, J. Fils, A. Clairon, N. Dimarcq, D. Holleville, P. Bouyer, and A. Landragin, J. Opt. B: Quantum and Class. Opt. \textbf{5}, S136-S142 (2003).

\bibitem{Leveque2009} T. L\'{e}v\`{e}que, W. Chaibi, A. Gauguet and A.
Landragin, \emph{in preparation}.

\bibitem{Santarelli1999} G. Santarelli, Ph. Laurent, P. Lemonde, A. Clairon, A.G. Mann, S. Chang, A.N. Luiten, C. Salomon, Phys. Rev. Lett. \textbf{82}, 4619 (1999).

\bibitem{Itano1993} W.M. Itano, J.C. Bergquist, J.J. Bollinger, J.M. Gilligan, D.J. Heinzen, F.L. Moore, M.G. Raizen, D.J. Wineland, Phys. Rev. A \textbf{47}, 3554 (1993).

\bibitem{mcguirk2002} J. M. McGuirk, G. T. Foster, J. B. Fixler, M. J. Snadden, and M.A. Kasevich, Phys. Rev. A \textbf{65}, 033608 (2002).

\bibitem{lamporesi2008} G. Lamporesi, A. Bertoldi, L. Cacciapuoti, M. Prevedelli, and G.M. Tino, Phys. Rev. Lett. \textbf{100}, 050801 (2008).

\bibitem{gravi2008} Similar results have been observed in a gravimeter experiment~\cite{legouet2008} while atomic clouds are not launched but only dropped.

\bibitem{weiss1994} D.S. Weiss, B.C. Young, and S. Chu, Appl. Phys. B \textbf{59}, 217-253 (1994).

\bibitem{moler1992} K. Moler, D.S. Weiss, M. Kasevich, and S. Chu,  Phys. Rev. A \textbf{45}, 342 (1992).

\bibitem{gauguet2008} A. Gauguet, T.E. Mehlst\"{a}ubler, T. L\'{e}v\`{e}que, J. Le Gou\"{e}t, W. Chaibi, B. Canuel, A. Clairon, F. Pereira Dos Santos and A. Landragin, Phys. Rev. A \textbf{78}, 043615 (2008).

\bibitem{fils2005} J. Fils, F. Leduc, P. Bouyer, D. Holleville, N. Dimarcq, A. Clairon, and A. Landragin, Eur. Phys. J. D \textbf{36}, 257 (2005).

\bibitem{Fils2002} J. Fils, PhD thesis, Universit\'{e} Paris XI (2002).

\bibitem{legouet2008} J. Le Gou\"{e}t, T.E. Mehlst\"{a}ubler, J. Kim, S. Merlet, A. Clairon, A. Landragin, and F. Pereira Dos Santos, Appl. Phys. B \textbf{92}, 133-144 (2008).

\bibitem{tidal} Robertson et al. Metrologia \textbf{38}, 71 (2001).

\bibitem{Dieckmann1998} K. Dieckmann, R.J.C. Spreeuw, M. Weidem\"{u}ller, and J.T.M. Walraven, Phys. Rev. A \textbf{58}, 3891 (1998).

\bibitem{nyman2006} R.A. Nyman, G. Varoquaux, F. Lienhart, D. Chambon, S. Boussen, J.-F. Clément, T. Müller, G. Santarelli, F. Pereira Dos Santos, A. Clairon, A. Bresson, A. Landragin, and P. Bouyer, Appl. Phys. B \textbf{84}, 673–681 (2006).

\bibitem{ertmer2009} W. Ertmer \emph{et al.}, Exp Astron \textbf{23}, 611–649 (2009).

\bibitem{giantRinglaser} K.U. Schreiber, J.-P. R. Wells and G. E. Stedman, Gen. relativ. Gravit. \textbf{40}, 935-943 (2008).

\bibitem{RFM2007} A. Landragin, B. Canuel, A. Gauguet, and P.
Tuckey, Revue Fran\c{c}aise de M\'{e}trologie \textbf{10}, 11-16
(2007).



\end{thebibliography}
\end{document}